\documentstyle[sprocl,epsf]{article}

\input{psfig}

\def\Journal#1#2#3#4{{#1} {\bf #2}, #3 (#4)}

\def\NCA{{\em Nuovo Cimento} A}
\def\NCC{{\em Nuovo Cimento} C}

\def\NIMA{{\em Nucl. Instrum. Methods} A}

\def\NPA{{\em Nucl. Phys.} A}

\def\PRC{{\em Phys. Rev.} C}

\def\ZPA{{\em Z. Phys.} A}

\def\be{\begin{equation}}
\def\ee{\end{equation}}
\def\bea{\begin{eqnarray}}
\def\eea{\end{eqnarray}}

\begin{document}

\title{FRAGMENTATION OF VERY HIGH ENERGY HEAVY IONS}

\author{M. GIORGINI$^{1,a}$ and S. MANZOOR$^{2,}$\footnote{for the EMU18 
Coll.: G. Giacomelli, M. Giorgini, H.A. Khan, G. Mandrioli, S. Manzoor, 
 L. Patrizii, V. Popa, I.E. Qureshi, M.A. Rana, M. Sajid, P. Serra, 
 M.I. Shahzad, G. Sher and V. Togo.}}

\address{$^1$ Dip. di Fisica and INFN, V. le C. Berti Pichat 6/2, I-40127 
Bologna, Italy\\E-mail: 
 miriam.giorgini@bo.infn.it \\~\\
$^2$ Radiation Physics Division, PINSTECH, P.O. 
Nilore, Islamabad, Pakistan,\\E-mail: 
 shahid.pins@dgcc.org.pk , manzoor@bo.infn.it}


\maketitle\abstracts{ {\bf Presented by M. Giorgini at the Int.
Conf. on Structure of the Nucleus
at the Dawn of the Century, Bologna (Italy), May 29-June 3, 2000.} \\~\\
A stack of CR39 (C$_{12}$H$_{18}$O$_7)_n$ nuclear track detectors
with a Cu target was exposed to a 158 A GeV lead ion beam at
the CERN-SPS, in order to study the fragmentation
properties of lead nuclei.
 Measurements of the total, break-up  and pick-up charge-changing
cross sections of ultrarelativistic Pb ions on Cu and CR39 targets
are presented and discussed.}

\section{Introduction}
\label{sec:introduction}
 We present experimental results on fragmentation charge-changing cross
sections of 158 A GeV lead ions (charge $Z=82e$) incident on Cu and 
CR39 targets. To detect and identify the relativistic ions, the 
nuclear track detector CR39 was used.
 When an ion crosses a nuclear track detector foil, it produces 
damages at the
level of molecular bonds, forming the so called ``latent track". During the
chemical etching of the detector in a basic water solution, etch-pit
cones are formed on both sides of the foil. The base area and the
height of each cone are functions of the Restricted Energy Loss (REL) of
the incident ion and thus of its charge $Z$~\cite{bb,ncim}. 

\section{Experimental procedure}\label{sec:experimental}
A stack made of CR39 nuclear track detectors with a Cu 
target was exposed in November 1996 at the CERN-SPS to a 
beam of 158 A GeV Pb ions.
 The exposure was performed at normal incidence. The total
number of lead ions incident on the stack was about $7.8  \times 10^4$,
 distributed in 8 spots. The central density in each spot was
around 1500 ions/cm$^2$.  \par
\begin{figure}[t]
\begin{center}
\mbox{\epsfysize=6.5cm
  \epsffile{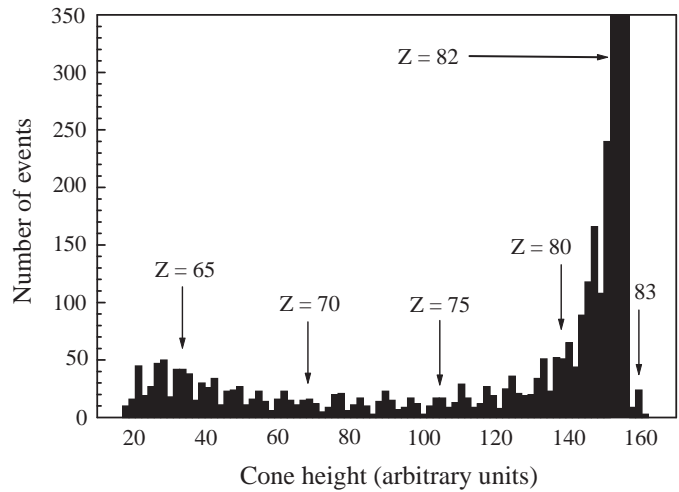}}
   \end{center}
\caption{Cone height distribution for Pb ions and heavy fragments measured
on one face of the CR39 sheet immediately after the Cu target.}
\label{fig:coni}
\end{figure}

 The stack had the following composition: 12 CR39 sheets 
$\sim 0.6$ mm thick, a Cu target
$\sim 10$ mm thick; 38 CR39 sheets $\sim 0.6$ mm thick. 
 In the present analysis, the CR39 sheets immediately before and after 
the Cu target and the last sheet of the stack were used.
 After exposure, the sheets were
 etched for 72 h in a 4N KOH water solution at a temperature of 
45 $^\circ$C. 
 Previous calibrations of the detectors have shown that 
for high $Z$ nuclei, the height of the etched cone is
more sensitive to $Z$ than its base area or diameter~\cite{nim}. 
 In order to separate the lead ions from 
the nuclear fragments with charge $Z \geq 63e$, we performed manual
  measurements of about 6300 cone heights using an optical Zeiss microscope
with a magnification of 40$\times$. Fig. \ref{fig:coni} shows the 
cone height distribution of Pb ions and heavy fragments measured 
on a single face of the CR39 sheet located after the Cu target.   
 The charge resolution obtained is about $0.2e$. 

\section{Total charge-changing cross sections}\label{sec:total}
Using the survival fraction of lead ions for the Cu and CR39 targets, we
measured the total charge-changing cross sections of lead ions
using the formula: 
\begin{equation}
\sigma_{tot}=\frac{A_T}{\rho_T t_T N_A} \ln {\frac{N_{in}}{N_{out}}} 
\label{eq:total}
\end{equation} 
where $N_{in}$ and $N_{out}$ are the numbers of lead ions
before and after the target, respectively; $N_A$ is Avogadro's number;
 $\rho_T,~A_T,~t_T$ are the density, the atomic 
mass and the thickness of the target. 
 The data are indicated by the black points in Fig. 2, the 
uncertainties are statistical only.  

\begin{center}
\begin{minipage}{.40\textwidth}
\begin{center}
\vspace{2mm}
\mbox{\hspace{-1cm} \epsfysize=7cm
  \epsffile{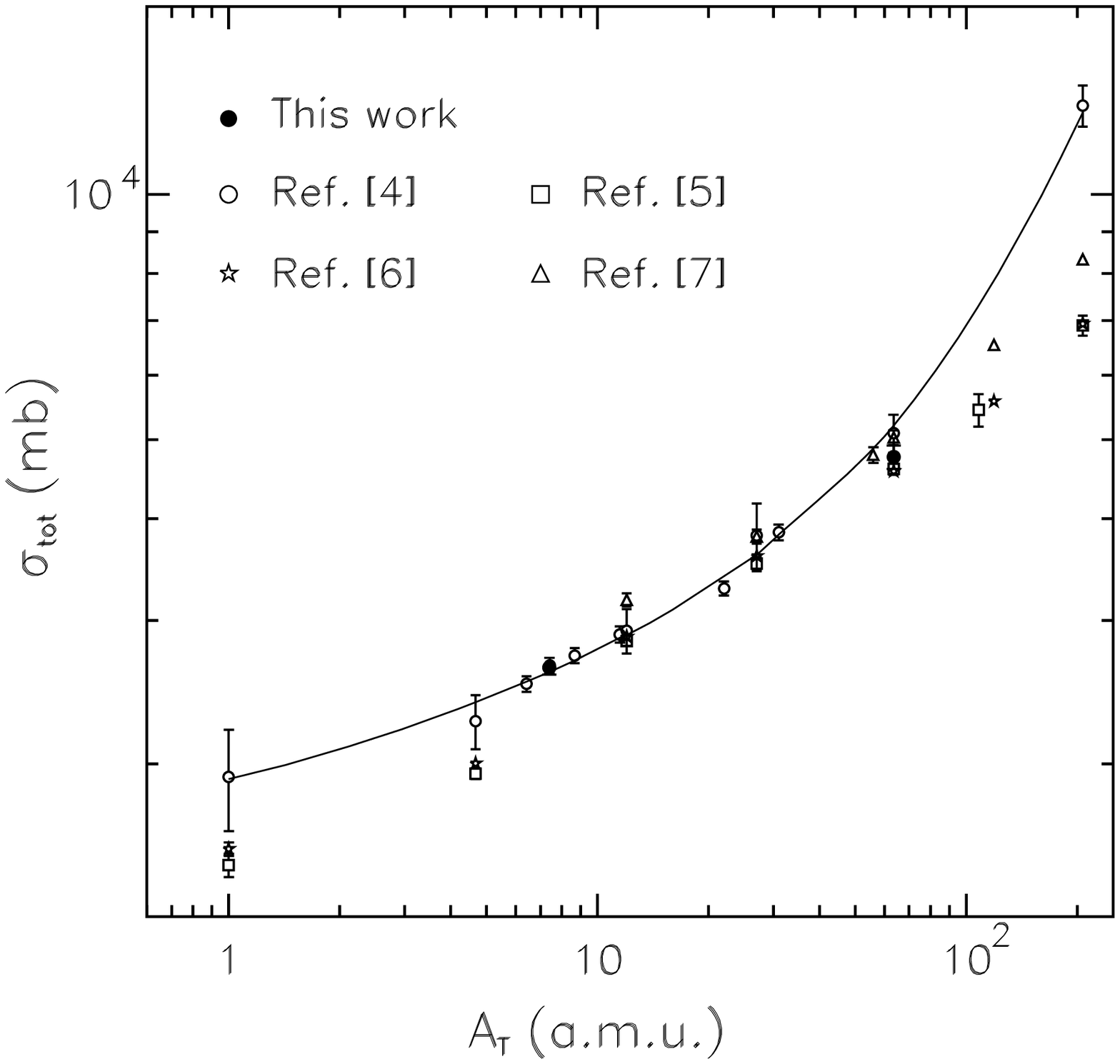}}
\end{center}
\vspace{2mm}
\end{minipage}\hfill
\begin{minipage}{.42\textwidth}
{\footnotesize Figure 2: Measured total fragmentation charge-changing
cross sections $\sigma_{tot}$ of
158 A GeV Pb projectiles versus the target mass number $A_T$: the black 
points are our data on Cu and CR39, the open points 
refer to data obtained by a similar experiment$^4$ using the same 
beam. The solid line represents the fit of all the data
to formula (2) of ref. [4]. 
 Results from a 10 A GeV Au beam incident on various 
targets~$^{[5-7]}$ are also shown.}
\end{minipage}
\end{center}

\setcounter{figure}{2}
As shown in Fig. 2, the
data are in agreement with previous data obtained by a similar experiment 
using the same beam and different targets with atomic masses ranging from 
4.7 a.m.u. (CH$_2$) to 207 a.m.u. (Pb)~\cite{sezioni}. 
The solid line in Fig. 2 is the fit of all the 
data to formula (2) of ref. [4] which yields $\chi^2/D.o.F.=0.7$. 
Results from other experiments using a 10 A GeV Au beam incident on
 various targets~$^{[5-7]}$ are also shown in Fig. 2.

\section{Partial fragmentation charge-changing cross sections}
\label{sec:partial}
\begin{figure}[t]
\begin{center}
\mbox{\epsfysize=5.9cm
  \epsffile{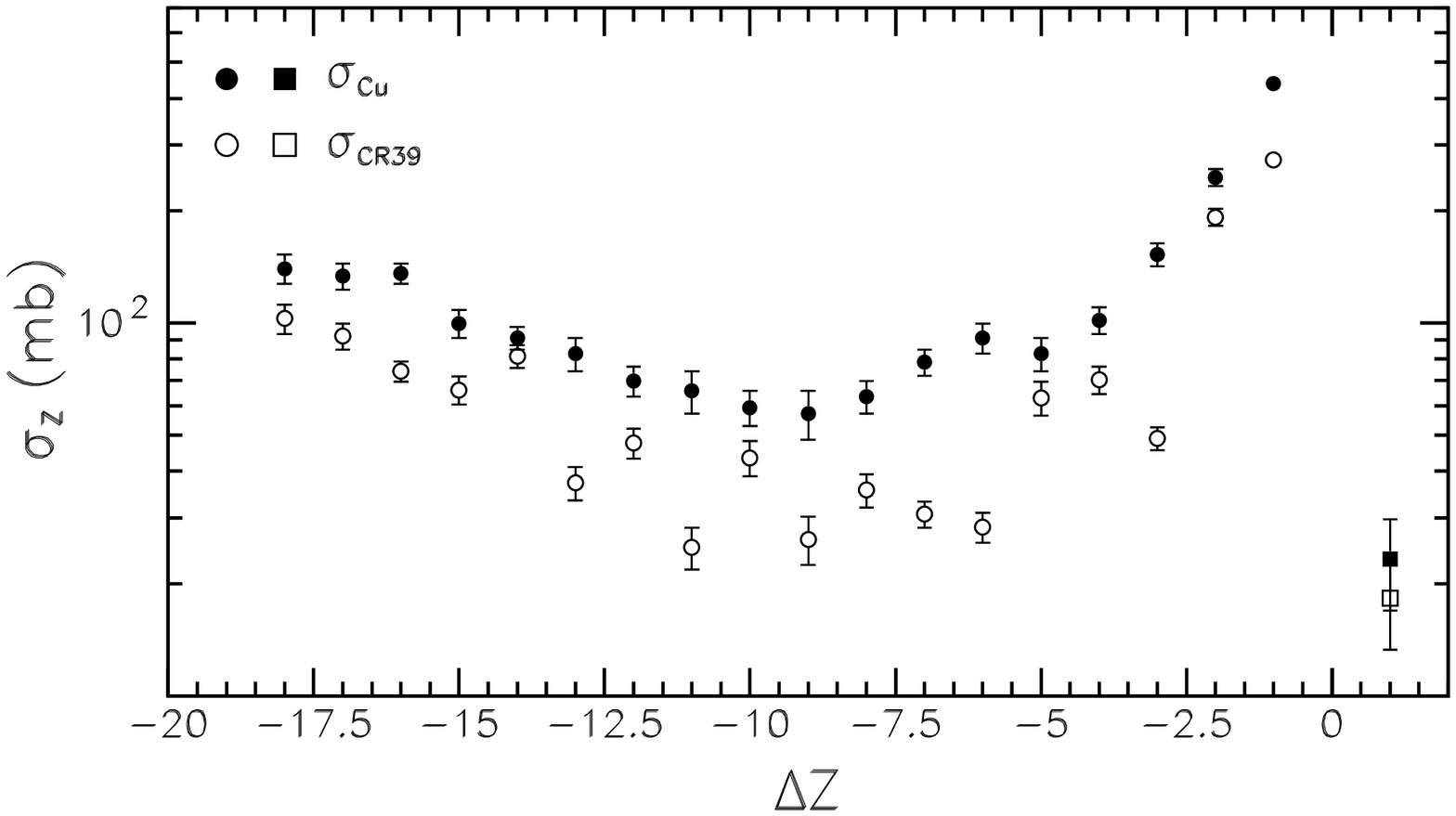}}
   \end{center}
\caption{Partial fragmentation charge-changing cross sections
for incident lead ions, $\sigma_{Z}$ versus $\Delta Z$ for the Cu 
and CR39 targets
computed with Eq. (\ref{eq:simple}). The black points refer to $\sigma_{Z}$
for Cu, the open points refer to $\sigma_{Z}$ for CR39. The squares 
refer to the pick-up cross sections. The errors are only statistical.}
\label{fig:partial}
\end{figure}

The partial fragmentation charge-changing cross sections of Pb ions yielding
fragments with charge $64e \leq Z < 82e$ were calculated for the
Cu and CR39 targets using the formula~\cite{bha}: 
\begin{equation}
\sigma_{Z}=\frac{A_T}{\rho_T t_T N_A}~{\frac{N_Z}{N_{82}}} \label{eq:simple}
\end{equation}
where $Z=64e \div 81e$,
  $N_Z$ is the number of fragment nuclei with charge $Z$ produced in the
target, $N_{82}$ is the number of unfragmented beam nuclei and
 $\rho_T,~A_T,~t_T,~N_A$ have the same meaning as in Eq. (\ref{eq:total}). 
 In this procedure, the successive fragmentation processes are
neglected. The results for the partial fragmentation cross sections of 
incident lead ions
on Cu and CR39 targets are shown versus $\Delta Z$ in Fig. \ref{fig:partial}.

The square points in Fig. \ref{fig:partial} refer to the charge pick-up 
cross sections, determined using Eq. (\ref {eq:simple}) where $N_Z$ 
is the number of nuclei with $Z=83e$ produced in the target.

\section*{Acknowledgments}
We thank the CERN SPS staff for the good performance
of the exposure, the technical staff of the Bologna INFN
and of the PINSTECH Laboratory.

\end{document}